\documentclass[conference,10pt]{IEEEtran}

\usepackage{algorithm,algorithmic}
\usepackage{amsmath,amssymb,mathrsfs}
\usepackage{array}
\usepackage{balance}
\usepackage{cite}
\usepackage{epsfig}
\usepackage{easyReview}
\usepackage{fancyhdr}
\usepackage{float}
\usepackage{glossaries}
\usepackage{graphicx,color}
\usepackage{hyperref}
\usepackage{mathtools}
\usepackage{multirow}
\usepackage{psfrag}
\usepackage{stfloats}
\usepackage{subfigure}
\usepackage{url}

\renewcommand{\eqref}[1]{(\ref{#1})}

\definecolor{sblue}{RGB}{0,51,120}
\definecolor{sred}{RGB}{139,0,139}
\definecolor{sg}{RGB}{46,139,87}

\makeglossaries

\ifCLASSINFOpdf
\else
\fi

\begin{document}
\title{\huge ResiTok: A Resilient Tokenization-Enabled Framework for Ultra-Low-Rate and Robust Image Transmission}
\author{ Zhenyu Liu, Yi Ma,  and Rahim Tafazolli\\
	{6GIC, Institute for Communication Systems, University of Surrey, Guildford, UK, GU2 7XH}\\
	{ Emails: (zhenyu.liu, y.ma, r.tafazolli)@surrey.ac.uk }
}

\markboth{}%
{Shell \MakeLowercase{\textit{et al.}}: Bare Demo of IEEEtran.cls for IEEE Journals}
\maketitle

\begin{abstract}
Real-time transmission of visual data over wireless networks remains highly challenging, even when leveraging advanced deep neural networks, particularly under severe channel conditions such as limited bandwidth and weak connectivity.  In this paper, we propose a novel Resilient Tokenization-Enabled (ResiTok) framework designed for ultra-low-rate image transmission that achieves exceptional robustness while maintaining high reconstruction quality. By reorganizing visual information into hierarchical token groups consisting of essential key tokens and supplementary detail tokens, ResiTok enables progressive encoding and graceful degradation of visual quality under constrained channel conditions. A key contribution is our resilient 1D tokenization method integrated with a specialized zero-out training strategy, which systematically simulates token loss during training, empowering the neural network to effectively compress and reconstruct images from incomplete token sets. Furthermore, 
 the channel-adaptive coding and modulation design dynamically allocates coding resources according to prevailing channel conditions, yielding superior semantic fidelity and structural consistency even at extremely low channel bandwidth ratios. Evaluation results demonstrate that ResiTok outperforms state-of-the-art methods in both semantic similarity and visual quality, with significant advantages under challenging channel conditions. 
\end{abstract}


\section{Introduction}
Real-time image and video transmission over wireless networks remains a significant challenge, especially in environments characterized by limited bandwidth and unreliable connections \cite{backgound1}. These conditions commonly arise during natural disasters, emergency situations in remote or rural areas, and mobile scenarios involving vehicles or satellites. Under such constraints, traditional methods combining image codecs (JPEG/JPEG2000/BPG) with advanced channel codes often fail \cite{backgound2}, resulting in severely distorted images. Even cutting-edge solutions like Apple's emergency message are limited to basic textual communication \cite{appleMessages2025}, highlighting the difficulty of transmitting richer media content over constrained channels.

Semantic communications (SemCom), empowered by artificial intelligence, offer a promising solution by extracting and transmitting semantic features of raw data, substantially reducing communication overhead \cite{jscc1,yang2025semantic,zhang2024MDJCM}. However, these approaches primarily focus on decreasing transmission cost under high-fidelity recovery constraints, limiting their performance in ultra-low-rate scenarios. Besides, a core limitation of current Joint Source-Channel Coding (JSCC) schemes \cite{jscc1,yang2025semantic,zhang2024MDJCM} is their reliance on a two-dimensional (2D) latent representation that preserves direct spatial mapping between latent tokens and image patches, restricting the full exploitation of inherent spatial redundancy for more efficient compression.

Recent breakthroughs in generative foundation models, such as Stable Diffusion \cite{Rombach_2022_CVPR} and GPT4o, have shown potential for ultra-low-rate semantic communication. By transmitting text descriptions with minimal visual sketch information, generative communication approaches \cite{xu2025generative} significantly reduce data rates while achieving notable semantic consistency. 
However, the computational complexity of foundation models introduces significant reconstruction latency, making them unsuitable for real-time applications and energy-limited devices. Furthermore, the separation of sketch information and prompt descriptions creates redundancy that reduces compression efficiency.

To address these challenges, the Transformer-based one-dimensional (1D) Tokenizer (TiTok) \cite{NEURIPS2024_e91bf7df} was recently proposed, providing superior compression efficiency by converting images into compact 1D discrete token sequences. Despite its remarkable compression capabilities, TiTok requires a fixed number of tokens for image recovery and is highly sensitive to transmission errors, limiting its practical deployment. Unlike classical image codecs, where errors typically result in localized artifacts, errors in tokenized representations propagate throughout the entire reconstruction, drastically degrading visual quality. Thus, there is an urgent need for tokenization-based communication systems capable of combining extreme compression with robust resilience to the channel instability.

In this paper, we propose ResiTok, a novel resilient tokenization-enabled framework for ultra-low-rate and robust image transmission. ResiTok transforms visual content into compact 1D token sequences and implements hierarchical importance-based token grouping, enabling progressive encoding with graceful quality degradation under adverse channel conditions. The main contributions are summarized as follows:
\begin{itemize}
    \item We propose a resilient tokenization-enabled framework for ultra-low-rate image transmission that preserves visual integrity under severe channel degradation while achieving superior compression efficiency.
    
    \item We introduce a novel 1D tokenization method integrated with a specialized zero-out training strategy to hierarchically organize tokens into essential ``key tokens" for basic reconstruction and supplementary ``detail tokens" for progressive visual enhancement.
    
    \item We include an adaptive transmission strategy that dynamically allocates resources based on channel conditions, optimizing the trade-off between bandwidth and resilience, complemented by a zero-padding scheme that enables coherent reconstruction using a single model.
    
    \item Experimental results validate that ResiTok outperforms existing state-of-the-art methods across diverse challenging environments, maintaining both semantic and visual consistency at significantly lower channel bandwidth ratios (down to 0.001).
\end{itemize}

\section{Resilient Tokenization Communication Framework}

\begin{figure*}[t]
    \centering
    \includegraphics[width=0.9\textwidth]{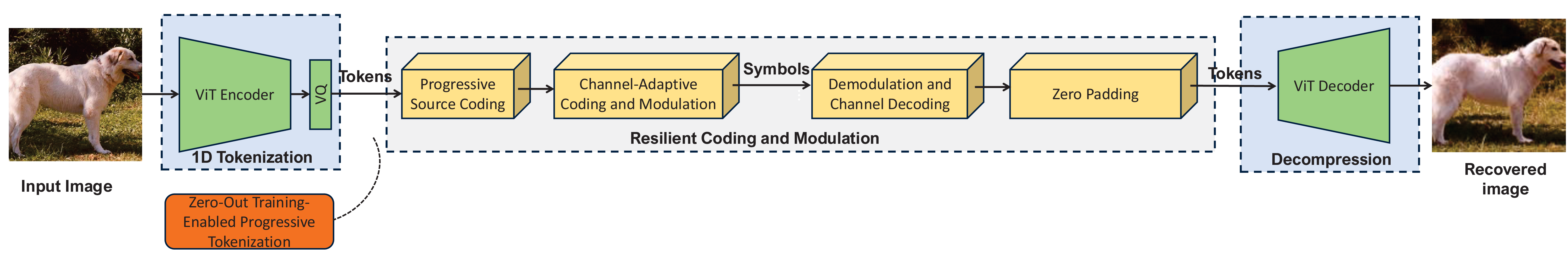}
     \vspace{-10pt}
    \caption{The proposed resilient tokenization communication framework for ultra-low-rate and robust image transmission.}
    \vspace{-10pt}
    \label{fig:framework}
\end{figure*}
    
Fig.~\ref{fig:framework} illustrates our proposed ResiTok framework for ultra-low-rate and robust image transmission. The framework consists of four primary functional blocks:

\begin{enumerate}
    \item \textbf{1D Tokenization:} The input image is processed through a Vision Transformer (ViT) encoder followed by Vector Quantization (VQ) \cite{NEURIPS2024_e91bf7df}, transforming visual content into a compact sequence of 1D tokens. 
    
    \item \textbf{Zero-Out Training:} This specialized methodology trains the neural model to organize and reconstruct visual content from incomplete token sets. During training, tokens are selectively masked to simulate transmission loss, enabling the system to maintain visual coherence even when significant portions of tokens are unavailable.
    
    \item \textbf{Resilient Coding and Modulation:} This block includes:
    \begin{itemize}
            \item \textit{Progressive Source Coding:} Organizes tokens into hierarchical groups—essential ``key tokens" that capture fundamental structural content, and ``detail tokens" that provide progressive refinement.
        
        \item \textit{Channel-Adaptive Coding and Modulation:} Dynamically allocates coding rates and modulation schemes based on channel conditions. 
        
        \item \textit{Demodulation and Channel Decoding:} Performs symbol demodulation and error correction to recover the transmitted token bitstream.
        
        \item \textit{Zero Padding:} Handles truncated or missing tokens by applying zero padding, enabling coherent image reconstruction despite incomplete information.
    \end{itemize}
    
    \item \textbf{Decompression:} The ViT decoder reconstructs visual content from the received tokens, generating a visually coherent output that preserves semantic meaning even with partial information.
\end{enumerate}

\section{Zero-Out Training Procedure}

\subsection{Tokenization Architecture}

Our framework builds on the tokenization architecture introduced in TiTok \cite{NEURIPS2024_e91bf7df}, which transforms images into compact token sequences with superior compression efficiency and reconstruction quality.

The 1D tokenization framework comprises three primary components:

\begin{enumerate}
    \item \textbf{Encoder:} A ViT encoder partitions the input image $\mathbf{x} \in \mathbb{R}^{H \times W \times 3}$ into patches and converts them into embedded vectors, where $H$ and $W$ represent the image height and width, respectively. The encoder introduces $N$ learnable latent tokens that capture essential information for image reconstruction, forming the latent representation $\mathbf{z}$.
    
    \item \textbf{Quantizer:} The latent representation $\mathbf{z}$ is discretized by a VQ, yielding a finite set of token representations:
    \begin{equation}
        \mathbf{y} = \text{Quant}(\mathbf{z}) = g_c(\mathbf{x}) \in \mathbb{R}^{N},
    \end{equation}
    where $g_c(\cdot)$ is the encoding function that converts the input image to tokens, and $N$ is the number of tokens.
    
    \item \textbf{Decoder:} In the de-tokenization phase, latent tokens $\mathbf{y}$ are dequantized and concatenated with mask tokens, then processed by the ViT decoder to generate the reconstructed image:
    \begin{equation}
        \hat{\mathbf{x}} = g_d(\mathbf{y}),
    \end{equation}
    where $g_d(\cdot)$ represents the decoding function that reconstructs the image from tokens.
\end{enumerate}

\subsection{Progressive Zero-Out Training Strategy}

Although TiTok provides excellent compression efficiency, it lacks inherent resilience to transmission errors. Our zero-out training strategy enhances the system's ability to preserve visual integrity under adverse channel conditions.

\begin{figure}[t]
    \centering
    \includegraphics[width=\columnwidth]{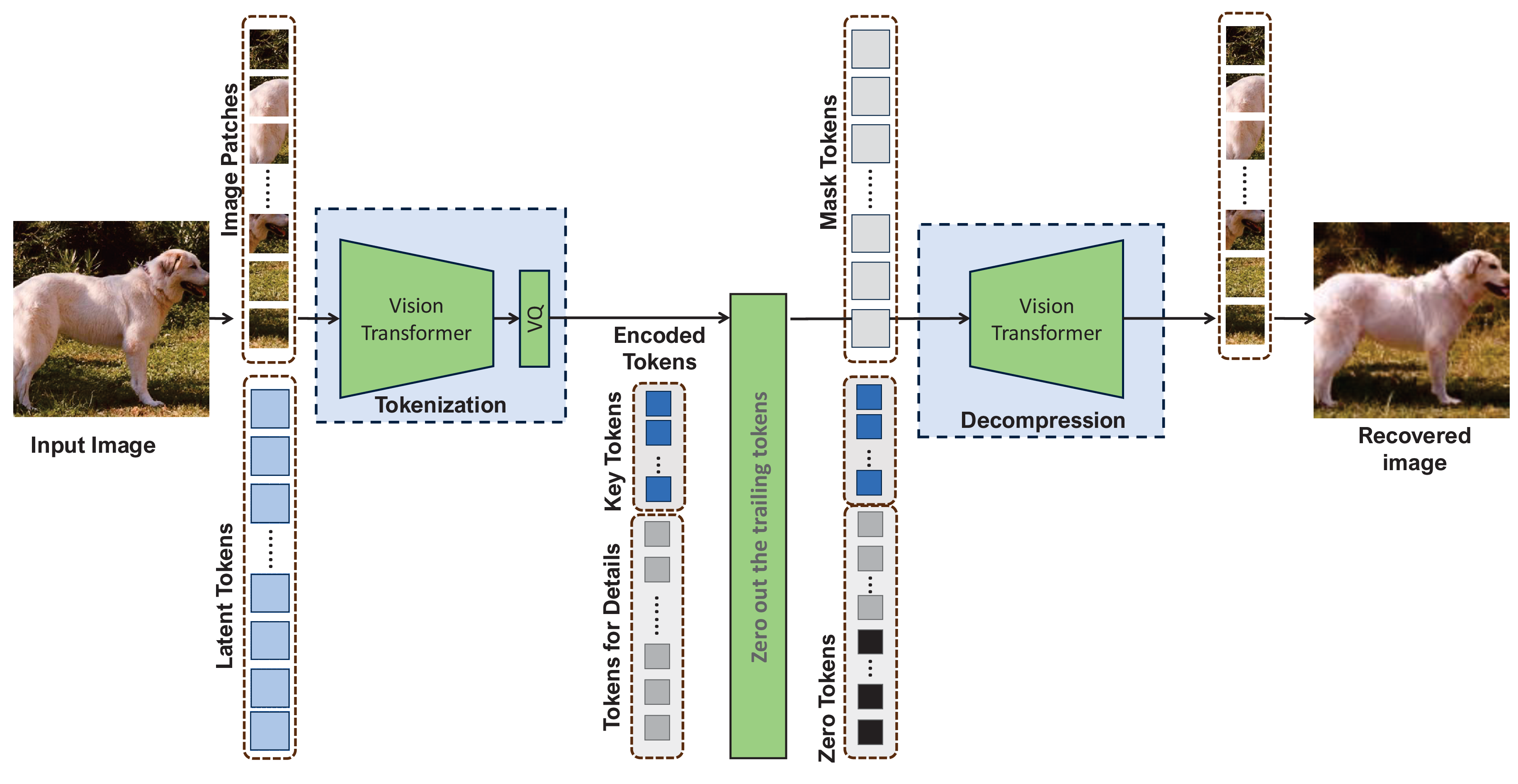}
    \caption{Training framework for progressive tokenization.}
    \vspace{-10pt}
    \label{figure_transnet}
\end{figure}

As illustrated in Fig.~\ref{figure_transnet}, our training methodology employs structured token masking to simulate partial transmission scenarios, consisting of three key steps:

\begin{enumerate}
    \item \textbf{Strategic Truncation Point Selection:} In each training iteration, a truncation point $t \in [|\mathbf{y}_K|, N]$ is selected from a uniform distribution: $t \sim U(|\mathbf{y}_K|, N)$,
    assigning equal probability to all valid truncation points and enabling training under various levels of token availability.
    
    \item \textbf{Selective Token Zeroing:} Based on the chosen truncation point, a masked token sequence is constructed:
    \begin{equation}
        \mathbf{y}_{\text{zeroing}} = \{\mathbf{y}_1, \mathbf{y}_2, \ldots, \mathbf{y}_t, \mathbf{0}_{t+1}, \mathbf{0}_{t+2}, \ldots, \mathbf{0}_N\},
    \end{equation}
    where $\mathbf{0}_j$ denotes a zero token that substitutes the original token at position $j$.
    
    \item \textbf{Resilience-Focused Training:} Using the two-stage training strategy in \cite{NEURIPS2024_e91bf7df}, the encoder generates hierarchical tokens including \emph{key tokens} for essential structural information and \emph{detail tokens} for refinement. The decoder is trained to reconstruct the image from the zeroed token sequence:
    \begin{equation}
        \hat{\mathbf{x}} = g_d(\mathbf{y}_{\text{zeroing}}).
    \end{equation}
\end{enumerate}

By forcing the model to reconstruct complete images from incomplete token sets, the network learns to encode semantically critical information in the earlier tokens while distributing refinement details across later tokens. This creates an information gradient across the token sequence, forming the resilience hierarchy essential for robust transmission. 

\section{Resilient Tokenization Enabled Ultra-Low-Rate and Robust Image Transmission}

Our system adopts a resilience-aware framework designed to transmit visual content efficiently while preserving robustness against channel variations. The following subsections describe the key components of the transmission pipeline.

\subsection{Tokenization Process}

An input image $\mathbf{x} \in \mathbb{R}^{H \times W \times 3}$ is processed to produce a sequence of tokens:
\begin{equation}
    \mathbf{y} = g_c(\mathbf{x}) \in \mathbb{R}^{N}.
\end{equation}

\subsection{Resilience-Aware Coding and Modulation}

\subsubsection{Progressive Source Coding}

The zero-out training enables the system to organize and reconstruct visual content from incomplete token sets. The tokens after 1D tokenization are classified into two categories:
\begin{enumerate}
    \item \textbf{Key Tokens} ($\mathbf{y}_K$): Contain fundamental information necessary for basic image reconstruction, prioritized during transmission.
    \item \textbf{Detail Tokens} ($\mathbf{y}_D$): Provide supplementary information for visual quality refinement, sorted in descending order of importance.
\end{enumerate}

The complete token set is expressed as:
\begin{equation}
    \mathbf{y} = \{\mathbf{y}_K, \mathbf{y}_D\}.
\end{equation}

Key tokens are always transmitted, while detail tokens can be truncated based on channel conditions, allowing receivers with limited capacity to decode essential parts while others can receive additional tokens for higher fidelity.

\subsubsection{Channel-Aware Coding and Modulation}

The system adapts transmission parameters based on channel conditions, where $\gamma$ represents the signal-to-noise ratio (SNR) of the communication channel. Under bandwidth constraints, key tokens receive guaranteed transmission, while detail tokens are selectively included based on available capacity:
\begin{equation}
    \mathbf{y}_{\text{transmitted}} = \mathbf{y}_K \cup \{ \mathbf{y}_{D,1}, \mathbf{y}_{D,2}, \ldots, \mathbf{y}_{D,n_t(\gamma)} \},
\end{equation}
where $n_t(\gamma)$ represents the maximum number of detail tokens that can be reliably transmitted under channel condition $\gamma$. Given the target channel bandwidth ratio (CBR), coding rate $r(\gamma)$, and modulation order $m(\gamma)$, we calculate $n_t(\gamma)$ as:
\begin{equation}
    n_t(\gamma) = \left\lfloor \frac{\text{CBR} \cdot H \times W \times 3 \cdot r(\gamma)}{b_{\text{token}} \cdot m(\gamma)} \right\rfloor - |\mathbf{y}_K|,
\end{equation}
where $b_{\text{token}}$ is the number of bits to represent a token.

The encoded bitstream and modulated signal are generated as:
\begin{align}
    \mathbf{c} &= C(\mathbf{y}_{\text{transmitted}}, r(\gamma)), \\
    \mathbf{s} &= M(\mathbf{c}, m(\gamma)),
\end{align}
where $C(\cdot)$ applies channel coding with rate $r(\gamma)$, and $M(\cdot)$ implements the modulation scheme with order $m(\gamma)$, both selected according to the adaptive modulation and coding standard \cite{3GPP_38_214}. The resulting modulated signal $\mathbf{s} \in \mathbb{C}^{N_s}$ has a channel bandwidth ratio of $\text{CBR} = \frac{N_s}{H\times W \times 3}$ relative to the original image dimensions.

The received signal $\mathbf{r}$ can be modeled as:
\begin{equation}
    \mathbf{r} = h \cdot \mathbf{s} + \mathbf{n},
\end{equation}
where $h$ represents channel gain and $\mathbf{n}$ denotes additive white Gaussian noise (AWGN).

\subsection{Reception and Reconstruction}

\subsubsection{Demodulation and Channel Decoding}

Upon reception, signals undergo demodulation and channel decoding:
\begin{align}
    \hat{\mathbf{c}} &= M^{-1}(\mathbf{r}, m(\gamma)), \\
    \hat{\mathbf{b}} &= C^{-1}(\hat{\mathbf{c}}, r(\gamma)),
\end{align}
where $\hat{\mathbf{c}}$ and $\hat{\mathbf{b}}$ denote the demodulated and decoded token sequence, and $M^{-1}(\cdot)$ and $C^{-1}(\cdot)$ represent the demodulation and decoding functions respectively.

\subsubsection{Token Reconstruction with Zero Padding}

For detail tokens not transmitted due to channel constraints, zero padding is applied:
\begin{equation}
    \hat{\mathbf{y}}_{D,i} =
    \begin{cases}
        \hat{\mathbf{y}}_{D,i}, & \text{if } i \leq n_t(\gamma), \\
        0, & \text{otherwise},
    \end{cases}
\end{equation}
with the complete reconstructed token set as:
\begin{equation}
    \hat{\mathbf{y}} = \mathbf{y}_K \cup \hat{\mathbf{y}}_D.
\end{equation}

\subsubsection{Image Reconstruction} 
Finally, the decoder receives the consistently structured token sequence (ensured by zero padding) and reconstructs the image:
\begin{equation}
    \hat{\mathbf{x}} = g_d(\hat{\mathbf{y}}),
\end{equation}
where $g_d(\cdot)$ denotes the decoding function.

\section{Performance Evaluations}
\label{sec:VI}
\subsection{Experimental Setup}
We implemented ResiTok and benchmark schemes using the PyTorch framework. All models are trained on the ImageNet dataset \cite{ILSVRC15}. During training, images are randomly cropped to patches of size $256 \times 256$. We adopt the ``TiTok-B" model parameters with a fixed number of tokens 256 for a 256 × 256 image patch, and follow the same training settings as in \cite{NEURIPS2024_e91bf7df}, with one key exception: due to hardware limitations (2 A6000 GPUs compared to 32 A100 GPUs in \cite{NEURIPS2024_e91bf7df}), our training is conducted for only 25\% of the epochs used in the original work. The codebook size of VQ is set to 4096, i.e., $b_\text{token} = 12$.

We evaluate the performance on two test datasets: the Kodak dataset \cite{kodakdataset}, with images of size $512 \times 768$, and the CLIC2021 test set \cite{clic2021}, which contains approximately 2K-resolution images.

For comparative evaluation, we incorporate several advanced methods. First, we include the most recent digital joint coding and modulation method (DJCM) proposed in \cite{zhang2024MDJCM} along with an enhanced backbone model from \cite{yang2025semantic}. Additionally, we consider the ``Prompt+Sketch" method \cite{xu2025generative}, which represents a state-of-the-art foundation model-assisted approach for generative image compressed transmission. Furthermore, we adopt the 1D tokenization model ``TiTok-S-128" (the shared model with the largest token length in \cite{NEURIPS2024_e91bf7df}) as an additional baseline. For ResiTok, TiTok and Prompt+Sketch, we choose the best-performing configuration of modulation and coding rate based on the AMC in the standard \cite{3GPP_38_214} under each SNR, and Turbo coding is used for channel coding due to its superior performance in low SNR and short code length scenarios \cite{channel_coding}. AWGN channel is employed for the performance evaluation.

\begin{figure}[t]
	\centering
	\includegraphics[width=\columnwidth]{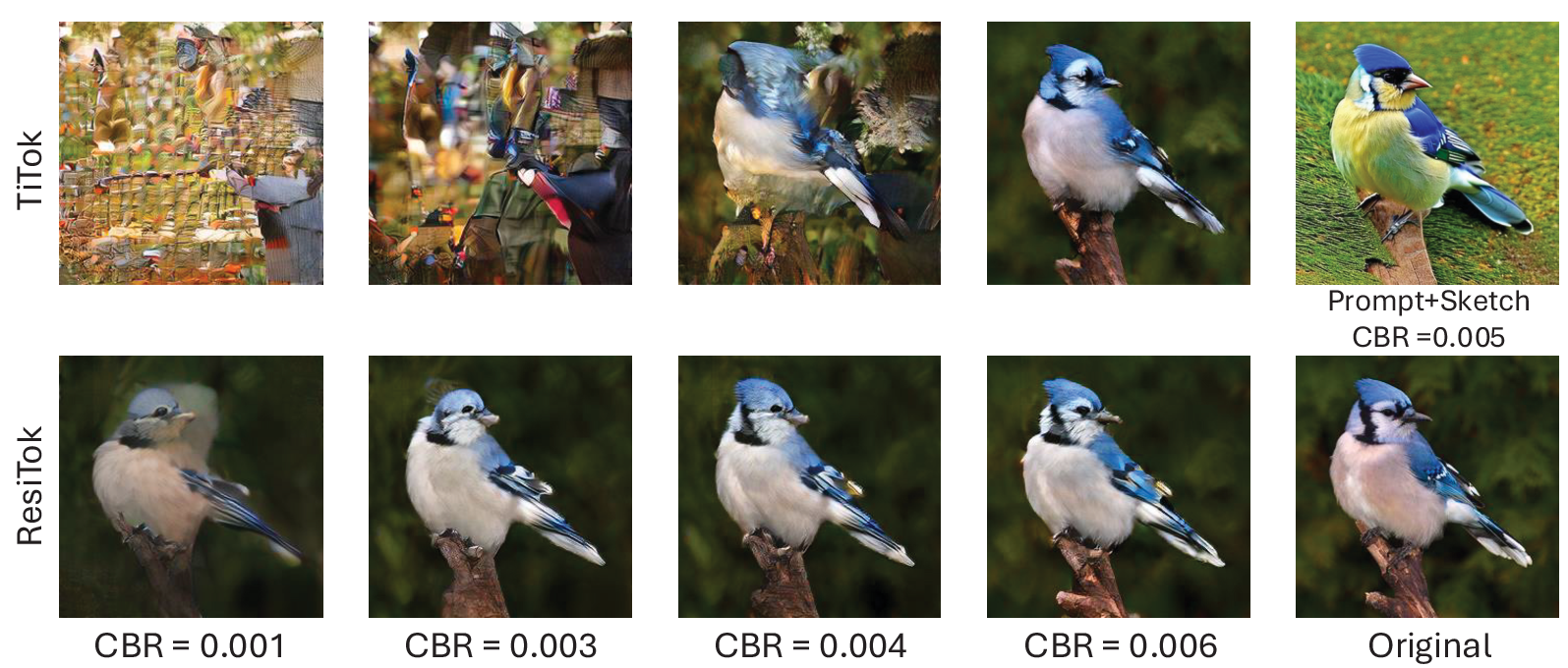}
    \vspace{-18pt}
	\caption{Visual quality and CBR comparison of the proposed approach against existing methods when SNR = 6dB using a 16-QAM modulation.}
	\label{fig:visual_demo}
    \vspace{-12pt}
\end{figure}

\subsection{Results Analysis}

\subsubsection{Visualization} Fig. \ref{fig:visual_demo} uses a bird as an example to compare our ResiTok framework with existing methods under challenging conditions (SNR = 6 dB, 16-QAM). The conventional TiTok method (top row) shows severe degradation—with catastrophic distortion and significant artifacts that render the bird unrecognizable at low channel bandwidth ratios (CBR). In contrast, our ResiTok method (bottom row) preserves basic structure and recognizability even at CBR = 0.001, with progressive quality improvements observed as the CBR increases to 0.006, approaching the original image quality. The Prompt+Sketch method (top-right) maintains reasonable semantic content at CBR = 0.005 but suffers from noticeable color and detail discrepancies. These results highlight the core strength of our progressive tokenization approach: it ensures graceful quality degradation and robust visual reconstruction under extreme bandwidth constraints.

\begin{figure}[tbp]
    \centering
    \subfigure[CLIP performance on CLIC2021]{
        \includegraphics[width=0.48\columnwidth]{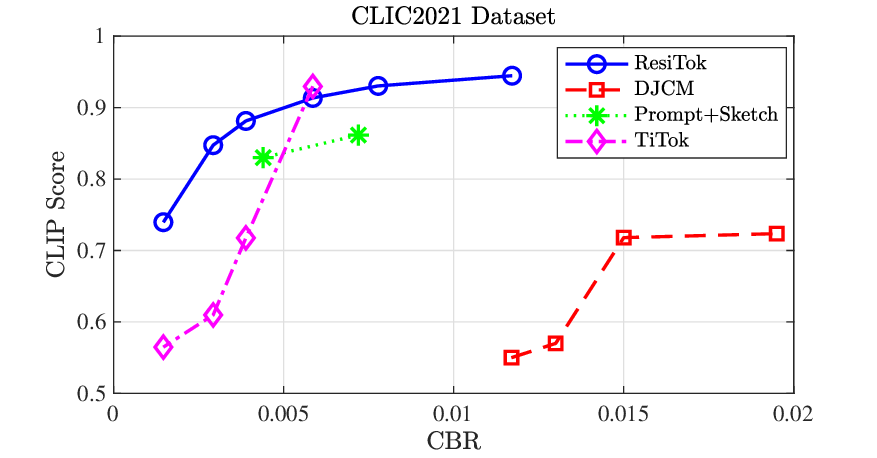}}
    \subfigure[PSNR performance on CLIC2021]{
        \includegraphics[width=0.48\columnwidth]{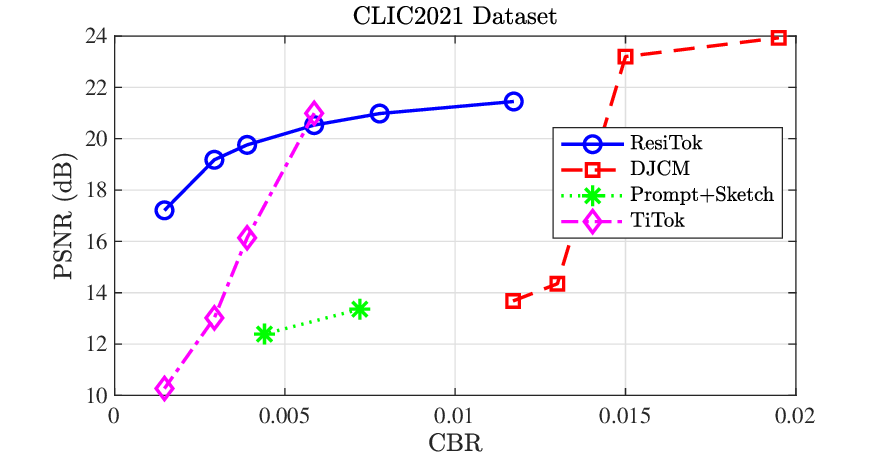}}
    
    \subfigure[CLIP performance on Kodak]{
        \includegraphics[width=0.48\columnwidth]{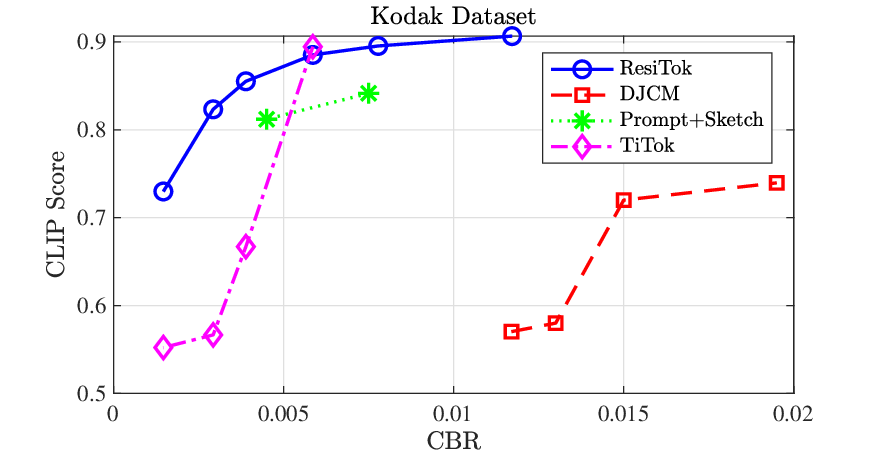}}
    \subfigure[PSNR performance on Kodak]{
        \includegraphics[width=0.48\columnwidth]{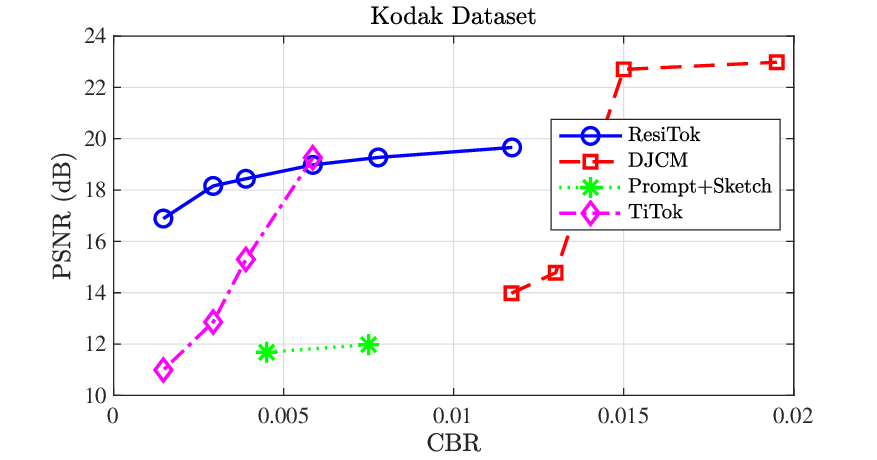}}
    \caption{Performance comparison versus CBR when SNR = 6dB: (a) CLIP score on CLIC2021 dataset, (b) PSNR on CLIC2021 dataset, (c) CLIP score on Kodak dataset, (d) PSNR on Kodak dataset.}
    \label{fig:cbr}
    \vspace{-10pt}
\end{figure}

\subsubsection{Performance comparison under different CBR}  Fig. \ref{fig:cbr} compares the performance of ResiTok against DJCM, Prompt+Sketch, and TiTok on the CLIC2021 and Kodak datasets using CLIP Score (semantic similarity) and PSNR (structural fidelity) when SNR = 6dB with a 16-QAM modulation. ResiTok consistently achieves higher CLIP Scores over the entire range of Channel Bandwidth Ratios (CBR), particularly excelling at very low bandwidths (CBR $<$ 0.005) by maintaining scores above 0.7—approximately 30\% higher than the nearest competitor at CBR = 0.001. 

In terms of PSNR, ResiTok outperforms competing methods in the low-to-medium bandwidth range (CBR $<$ 0.012); for example, at CBR = 0.003, ResiTok attains about 19dB while TiTok reaches only 13dB. Because we only trained a single model with the maximum token length = 256, ResiTok cannot achieve CBR $\geq$ 0.012 yet.
Meanwhile, the conventional TiTok method shows significant improvement with increasing bandwidth but remains inferior at extremely low CBR. The Prompt+Sketch method offers moderate semantic preservation but consistently lower PSNR compared to ResiTok and TiTok.
These consistent performance trends across both datasets validate ResiTok's exceptional ability to maintain semantic and structural integrity under severely constrained channel conditions.

\begin{figure}[!htb]
    \centering 
    \subfigure[CLIP Score]{
    \includegraphics[width=0.43\textwidth]{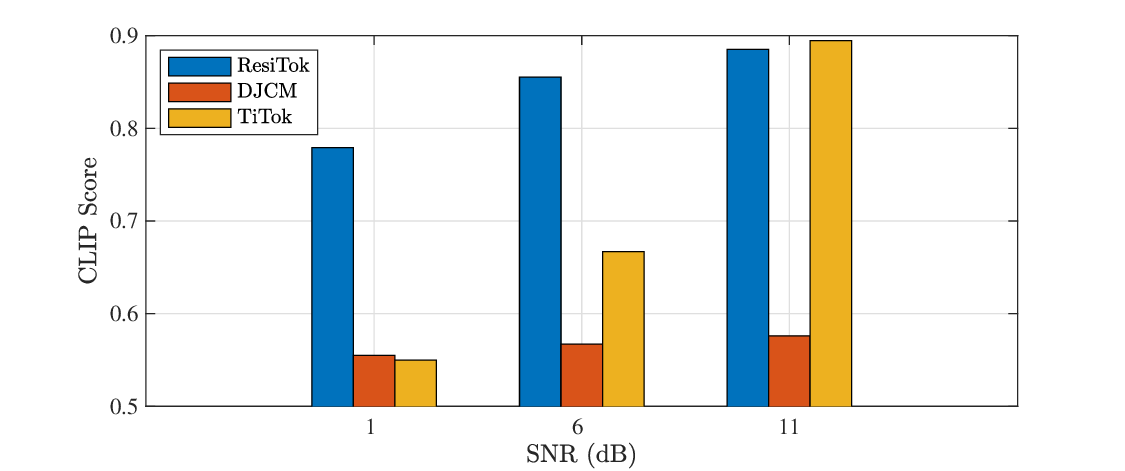} 
    } 
    \vspace{-6pt}
    \subfigure[PSNR]{
    \includegraphics[width=0.43\textwidth]{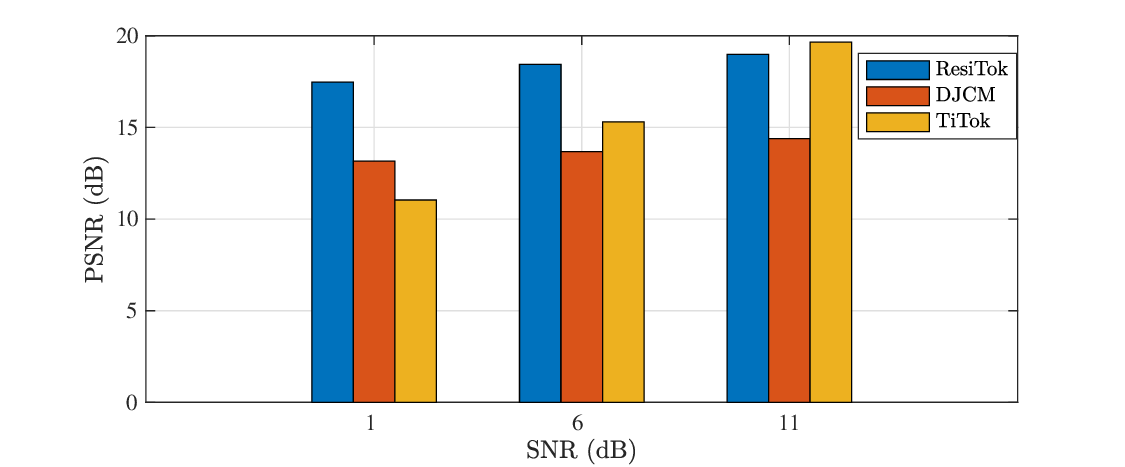}
    } 
    \caption{Performance versus SNR over Kodak dataset when CBR = $\frac{1}{256}$: (a) CLIP Score and (b) PSNR metrics.}
    \label{fig:SNR}
\end{figure}

\subsubsection{Performance comparison under different SNR}  Fig. \ref{fig:SNR} compares ResiTok with DJCM and TiTok under an ultra-low CBR =$\frac{1}{256}$ across three SNR levels: severely degraded (1 dB), challenging (6 dB), and favorable (11 dB). For semantic preservation, ResiTok maintains a high CLIP score of about 0.78 at 1 dB—approximately 42\% higher than its competitors—and improves to approximately 0.85 at 6 dB. At 11 dB, TiTok slightly outperforms ResiTok, indicating that conventional tokenization works well under ideal conditions. 

In terms of visual quality, ResiTok achieves approximately 17.5 dB PSNR at 1 dB and 18.5 dB at 6 dB, significantly exceeding the performance of both DJCM and TiTok under poor to moderate conditions. At 11 dB, TiTok reaches approximately 0.6 dB better PSNR, yet its rapid degradation under adverse conditions limits its overall reliability. These results validate that ResiTok delivers robust semantic and structural fidelity, ensuring reliable image transmission even under severely constrained channel conditions.

\subsection{Computational Complexity Analysis}

\begin{table}[t]
\begin{center}
\caption{Computational complexity comparison}
\vspace{-3pt}
\label{tab:complexity}
\renewcommand\arraystretch{1.5}
\begin{tabular}{|c|c|c|c|c|}
\hline
\textbf{Metric} & \textbf{DJCM} & \textbf{TiTok} & \textbf{Prompt+Sketch} & \textbf{ResiTok} \\
\hline
Parameters & 37.2 M & 84.4 M & 1773.9 M & 205.5 M \\
\hline
FLOPs & 104.3 G & 222.1 G & 24.6 T & 304.6 G \\
\hline
\end{tabular}
\vspace{-16pt}
\end{center}
\end{table}

Table \ref{tab:complexity} summarizes the computational complexity of the compared methods. The ResiTok framework, employing a base model with 205.5 M parameters, represents a 2.4$\times$ increase relative to TiTok-S-128 (84.4 M) due to the larger model capacity required for progressive encoding and enhanced resilience. Although DJCM is the most lightweight with only 37.2 M parameters, its performance degrades significantly at low bandwidth ratios. The Prompt+Sketch method is substantially more resource-intensive, using 1773.9 M parameters—approximately 8.6$\times$ more than ResiTok—owing to its reliance on large-scale generative models.

Regarding FLOPs, ResiTok requires 304.6 G, which is 37\% higher than TiTok's 222.1 G FLOPs. This modest increase in computational cost is justified by the substantial gains in transmission resilience. In contrast, DJCM has the lowest computational load (104.3 G FLOPs) but underperforms in challenging channel conditions. Furthermore, Prompt+Sketch demands an impractical 24.6 T FLOPs—over 80 times more than ResiTok— making it unsuitable for real-time applications. These comparisons illustrate that ResiTok achieves an optimal trade-off between computational efficiency and transmission robustness, making it well-suited for deployment in real-world, bandwidth-constrained environments.

\section{Conclusion}
This paper has presented the ResiTok, a novel resilient tokenization-enabled framework for ultra-low-rate and robust image transmission.  By integrating 1D tokenization with our specialized zero-out training procedure, ResiTok organizes visual data into hierarchical token groups—essential key tokens and supplementary detail tokens—enabling progressive encoding that gracefully adapts to varying channel conditions. The system preserves critical semantic content even under severe bandwidth constraints while allowing quality refinement when conditions improve. Extensive evaluations on the Kodak and CLIC2021 datasets demonstrate that ResiTok significantly outperforms state-of-the-art methods in semantic similarity and visual fidelity at ultra-low channel bandwidth ratios.

\bibliographystyle{IEEEtran}
\bibliography{SemRef} 
	
\end{document}